\documentclass[preprint,5p]{elsarticle}
\usepackage{graphicx}
\usepackage{dcolumn}
\usepackage{bm}
\usepackage{hyperref}
\usepackage{amssymb}
\usepackage{hyperref}
\RequirePackage[normalem]{ulem} 
\RequirePackage{color}\definecolor{RED}{rgb}{1,0,0}\definecolor{BLUE}{rgb}{0,0,1} 

\begin{document}
\newcommand{\be}{\begin{equation}}
\newcommand{\ee}{\end{equation}}
\newcommand{\ba}{\begin{eqnarray}}
\newcommand{\ea}{\end{eqnarray}}
\newcommand{\Gam}{\Gamma[\varphi]}
\newcommand{\Gamm}{\Gamma[\varphi,\Theta]}
\thispagestyle{empty}

\title{Quantum entanglement  of two harmonically trapped dipolar particles
}

\author{Przemys\l aw Ko\'scik,
Institute of Physics,  Jan Kochanowski University\\
ul. \'Swi\c{e}tokrzyska 15, 25-406 Kielce, Poland}



\begin{abstract}
We study systems  of two identical dipolar particles confined in
quasi one-dimensional harmonic traps. Numerical results for the
dependencies of the entanglement on the control parameters of the
systems  are provided and discussed in detail. In the  limit of a
strong interaction between the particles, the occupancies and the
von Neumann entropies of the
 bosonic and fermionic ground states are derived in closed analytic forms
 by applying the
 harmonic approximation. The strong correlation regimes
of the  system with the dipolar bosons  and  the  system with  the
charged ones are compared with each other in regard to aspects of
their entanglement.
\end{abstract}
\maketitle


\section{Introduction} In recent years, the study of  quasi one-dimensional (1D) systems  of cold atoms
with a short-range interaction has drawn considerable attention. In
particular, an observation of the Tonks--Girardeau (TG) systems
\cite{kin}, where bosonic systems behave as    gases of spinless
non-interacting fermions \cite{fer}, has inspired a great interest
in exploring their properties.
 The recent technological developments
have also opened up perspectives for  the experimental realization
of systems of spatially confined particles with long-range
dipole--dipole  interactions (DDI)\cite{ex1,ex2,ex3,ex4,ex5},
providing, among other things, new possibilities for studying
quantum correlation effects in many-body systems.
 Since then,
there has been a remarkable increase of  interest in understanding
the properties of such systems
\cite{dip0,dipl,dip2,dip3,dip4,dip4.1}. For  recent developments in
studies of ultra-cold gases, including   dipolar quantum gases, see
the overview \cite{dip5}.

Also, the  study of  entanglement has attracted much attention
within the last few years. Particularly the research activity has
 expanded towards investigating the entanglement properties in
various systems composed of interacting particles.
 For instance, the
recent studies include model systems such as the Moshinsky atom
\cite{1,2,3,nagy}, helium atoms and helium-like atoms
 \cite{4,7,7.1}, quantum dot systems \cite{5,6,9.1,ghjd}, and
1D systems of atoms interacting via  a short-range contact
interaction \cite{cont0,cont1,kos10}. For details concerning the
recent progress in entanglement studies of quantum composite
systems, see \cite{tihy}. However, according to our best knowledge,
studies on the entanglement of dipolar particles have not yet drawn
much attention. In this paper, we address this issue and gain some
insight into the quantum entanglement properties of systems composed
of two particles confined in a harmonic trap

\begin{eqnarray} V(\textbf{r})={m\over 2} (\omega^2
x^2+\omega_{\bot}^2\rho^2),\end{eqnarray} with the DDI
  modeled by
\begin{eqnarray} U(\textbf{r})={d^2\over |\textbf{r}|^3}(1-3 \mbox{cos}^2 \theta_{rd}),\end{eqnarray}
where $\mbox{cos}^2 \theta_{rd}={\vec{r}\cdot \vec{d}/ rd}$, and
$d^2$ being the strength of the DDI. For the sake of simplicity, we
 focus here on the regime of strong anisotropy,  $\epsilon=\omega_{\bot}/\omega>>1$.  The
theoretical description of such  quasi-1D systems can be simplified,
since one can assume that the particles stay in the lowest
transverse confinement mode and the single-mode approximation (SMA)
can be
 applied.   Within this approximation, the system of
four dipolar bosons in a trap of anisotropy of
 $\epsilon=50$ has recently been  considered in \cite{dipl}, wherein   the effects of  the interaction strength on various
characteristics (such as the density, the momentum distribution, and
the occupation number distribution) have been determined.

For the  two-particle system under consideration here, the
Hamiltonian in the SMA
 is
\begin{equation}
H=\sum_{i=1}^{2}[-{1\over 2}{\partial^2\over \partial
x_{i}^2}+{1\over 2} x_{i}^2]
 + g U(|x_{2}-x_{1}|).\label{EOM11}
\end{equation}
with  \be U(x)=\epsilon^{3\over 2}(2 \sqrt{\epsilon}x-e^{-{\epsilon
x^2\over 2}}\sqrt{2\pi}(1+\epsilon x^2)
\mbox{erfc}({\sqrt{\epsilon\over 2}}x)),\ee
 where we have omitted the
short-range contact interaction in order  to explore only the pure
dipolar effects (for details on this point, see, for example,
\cite{dipl}). The coordinates and the energies are measured in terms
of $\sqrt{{m\omega}/{\hbar}}$, ${\hbar \omega}$, respectively, and
the dimensionless coupling $g$   is related to the control
parameters of the system by $g={d^2 \sqrt{\omega} m^{3\over
2}(1+3\mbox{cos}2\theta)/ 8\hbar^{5\over 2}}$.

The system (\ref{EOM11})  has the convenient feature that the
center of mass (cm.) and relative motion can be decoupled.  In terms
of
$$ x={{x}_{2}-{x}_{1}\over \sqrt{2}}, X={x_{1}+x_{2}\over
\sqrt{2}},$$
 the Hamiltonian  (\ref{EOM11}) separates into $H= H^{x} +H^{X}$,
where the cm. Hamiltonian  $H^{X}=-{1/ 2}d^2_{X} +X^2/2$
 is exactly solvable and the relative motion Hamiltonian is given by
\begin{equation}
H^{x}=-{1\over 2}{d^2\over d x^2}+{1\over 2} x^2
 + g U(\sqrt{2}|x|) .\label{EOM1}
\end{equation}
The Taylor expansion of  $U(\sqrt{2}|x|)$ around  $\epsilon=\infty$
gives ${ \sqrt{2} |x|^{-3}}$, so that  the relative motion
 is governed in the strictly 1D
limit by
\begin{equation}
H^{\epsilon\rightarrow\infty}= -{1\over 2}{d^2\over d x^2}+{1\over
2} x^2
 + {g \sqrt{2}\over |x|^3}.\label{EOM2}
\end{equation}

In this paper, we  discuss  the effect  of both the anisotropy
parameter $\epsilon$ and the interaction strength $g$ on the
entanglement in the bosonic and fermionic ground states. Moreover,
we derive, within the framework of the harmonic approximation (HA),
closed form expressions for the natural orbitals and their
occupancies in the $g\rightarrow\infty$ limit. Furthermore, using
these results, we analytically calculate the corresponding
asymptotic bosonic and fermionic von Neumann (vN) entropies and
compare their values with the ones  obtained numerically.

This paper is structured as follows. In Section \ref{section2}, we
discuss the entanglement characteristics of quasi-1D systems of two
spinless identical particles. Section \ref{section3} is devoted to
the results, and some concluding remarks are made in Section
\ref{sum}.

\section{The measure of entanglement }\label{section2}
The tool that is usually used  to  characterize quantum entanglement
is the spectrum of the one-particle reduced density matrix (RDM)
\cite{vn1}. For systems of a quasi-1D geometry composed of two
spinless particles, the RDM takes the form \be
\rho(x,x^{'})=\int_{-\infty}^{\infty} \psi
(x,y)\psi(x^{'},y)dy\label{ded}.\ee  The eigenvectors and
eigenvalues of the RDM are sometimes called the natural orbitals and
the occupancies, respectively, and we shall do the same.  The
normalization conditions for the occupancies of the bosonic $(+)$
and fermionic $(-)$ states give $\sum_{l}\lambda_{l}^{(+)}=1$, and
$2\sum_{l}\lambda_{l}^{(-)}=1$, where the factor $2$ comes from the
fact  that the occupancies  of the fermionic state are doubly
degenerate. As discussed in \cite{gh}, the state of two identical
bosons is non-entangled only in two cases, i.e., when its Schmidt
number ($\mbox{Sn}$) is  $1$ or $2$, which corresponds to
$\lambda_{i}^{(+)}=1$ and to
$\lambda_{i}^{(+)}=\lambda_{j}^{(+)}=0.5$, respectively. As to a
fermion state, it is non-entangled if, and only if, its total
wavefunction can be expressed as one single determinant \cite{gh},
which in turn corresponds to $\lambda_{i}^{(-)}=0.5$.

We will measure the amount of the entanglement via the vN entropy,
\be \mbox{S}_{vN}=\mbox{S}+\mbox{S}_{0},\label{vbn}\ee where
$\mbox{S}=-\mbox{Tr}[\rho \mbox{Log}_{2} \rho]$ is the ordinary vN
entropy \cite{vn1} and $\mbox{S}_{0}=0$ and $\mbox{S}_{0}=-1$ stand
for the bosonic ($B$) and fermionic ($F$) states, respectively. In
terms of the occupancies,
 we have
  $ \mbox{S}_{B}=-\sum_{l=0}
\lambda_{l}^{(+)} \mbox{log}_{2} \lambda_{l}^{(+)}$, $
\mbox{S}_{F}=-1-2\sum_{l=0} \lambda_{l}^{(-)} \mbox{log}_{2}
\lambda_{l}^{(-)}$.   The  measure (\ref{vbn}) vanishes for the non-entangled
points discussed above, except for the bosonic   states
with $\mbox{Sn}=2$, ($\mbox{S}_{B}=1$).  As  noted in the already cited
\cite{gh}, the vN entropy alone is insufficient to
distinguish whether the bosonic state is entangled or not, since
the following may happen: $\mbox{S}_{B}=1$ for the state with
$\mbox{Sn}$ different than $2$. For example, such a situation was
observed in the system of bosons interacting via the short-range
contact potential confined in a split trap \cite{cont0}.

\section{Results and discussion}\label{section3}

\subsection{The  weak-interaction limit}
 In the  limit as $\epsilon\rightarrow \infty$ (the strictly 1D case) the system  of bosons
gets fermionized for any $g\neq 0$ \cite{fer}, which is due to the
singular behavior of the interaction potential
$|x_{2}-x_{1}|^{-{3}}$ at $x_{1}=x_{2}$. In this limit, the two
 bosonic ground-state wavefunctions approach, as  ${g\rightarrow0}$, the modulus of
the Slater determinant
$\psi^{g\rightarrow0}_{B}(x_{1},x_{2})=2^{-{1/
{2}}}|\mbox{det}_{n=0,j=1}^{1,2}(\varphi_{n}(x_{j}))|$, where
$\varphi_{n}$ are the single-particle orbitals of the ideal system
($g = 0$). The above function is nothing else but a TG wavefunction,
which means  that  the 1D dipolar bosons form  a TG gas in the weak
interaction regime. We have determined the value of the vN entropy
associated with $\psi^{g\rightarrow0}_{B}$ by calculating the
occupancies numerically
  through a discretization technique (see,
for example, \cite{9.1}). The value obtained by us
$\mbox{S}^{g\rightarrow0}_{B}\approx0.9851$ agrees well  with that
reported in \cite{cont1} ($0.984$). On the other hand, in the
non-interacting case  $g=0$, the bosonic and fermionic ground-state
wavefunctions are given by a simple product
$\psi^{g=0}_{B}(x_{1},x_{2})=\varphi_{0}(x_{1})\varphi_{0}(x_{2})$
and a  Slater determinant
 $\psi^{g=0}_{F}(x_{1},x_{2})=2^{-{1/
{2}}}\mbox{det}_{n=0,j=1}^{1,2}(\varphi_{n}(x_{j}))$, respectively,
and
 these states must  be
regarded as non-entangled (their corresponding vN entropies vanish).
One can importantly conclude that
  the vN entropy  of the fermionic ground-state is
always continuous at $g=0$, which is in contrast
 to the vN entropy of the bosonic ground-state,  which  exhibits  a discontinuity at
this point as $\epsilon\rightarrow\infty$.

\subsection{The  strong-interaction limit }\label{dd}

Now we come to the point where we explore  the limit of
$g\rightarrow\infty$ by use of
 a scheme   developed in
 \cite{9.1}. Due to the long-range nature of the DDI, the larger is $g$, the
larger is the average distance between the particles. Hence, and
because of the fact that the Taylor expansion of $U(\sqrt{2}|x|)$
around $|x|=\infty$ gives ${ \sqrt{2} |x|^{-3}}$, one can conclude
that at very large $g$ the distance between the particles is large
enough so that  the interaction among them does not depend on
$\epsilon$. Thus, as long as we are   interested in the regime when
$g\rightarrow\infty$, we can focus only on   Eq. (\ref{EOM2}).

To begin with, we expand the  relative motion potential in
(\ref{EOM2}), $V(x)=x^2/2+{g\sqrt{2} |x|^{-3}}$, around its local
minimum $ x_{c}=2^{1/ 10}( 3 g)^{1/ 5}$, retaining only the terms up
to second order, $V(x)\approx V(x_{c})+{5} (x - x_{c})^2/2$. The
relative motion Schr\"{o}dinger equation with such a potential is of
course exactly solvable and approximations to the lowest even ($+$)
and odd ($-$)
 wavefunctions  can be  constructed as follows \be
\psi_{rel}^{(\pm)}=C^{(\pm)}(g)(e ^{-{\sqrt{5}\over
2}(x-x_{c})^2}\pm e ^{-{\sqrt{5}\over 2}(x+x_{c})^2}), \ee where
$C^{(\pm)}(g)$ is the normalization factor that tends to ${5^{1\over
8}/ \sqrt{2}{\pi}^{1\over 4} }$ as $g\rightarrow\infty$ both in the
$(+)$ and $(-)$ case. After taking into account the cm. ground-state
wavefunction, \be
\psi^{(\pm)}(x_{1},x_{2})=\psi_{rel}^{(\pm)}(x){e^{-{X^2\over
2}}\over \pi^{1\over 4}},\nonumber\ee and then changing the
variables back to $x_{1}$ and $x_{2}$, we obtain final forms of the
approximations to the lowest total
 symmetric $(+)$ and antisymmetric $(-)$ wavefunctions as
  \be
\psi^{(\pm)}(x_{1},x_{2})=
\zeta(x_{1},x_{2})\pm\zeta(x_{2},x_{1})\label{ex},\ee with \be
\zeta(x_{1},x_{2})={C^{(\pm)}(g)\over \pi^{1\over 4}}e^{-{1\over
4}(x_{1}+x_{2})^2-{\sqrt{5}\over 2}({x_{2}-x_{1}\over
\sqrt{2}}-x_{c})^2}.\label{ddd}\ee Subsequently we
 translate   the coordinates by $ x_{1}\mapsto
\tilde{x}_{1}-{x_{c}/ \sqrt{2}}, x_{2}\mapsto \tilde{x}_{2}+{x_{c}/
\sqrt{2}}$, which turns   Eq. (\ref{ddd}) into $x_{c}$-independent
form: \be \tilde{\zeta}(\tilde{x}_{1},\tilde{x}_{2})=
{C^{(\pm)}(g)\over \pi^{1\over 4}} e^{-{1\over
4}(\sqrt{5}(\tilde{x}_{2}-\tilde{x}_{1})^2+(\tilde{x}_{1}+\tilde{x}_{2})^2)}
\label{polo}. \ee Due to the fact that the above function
 is symmetric under permutations of coordinates, it
always has a Schmidt decomposition in the form \cite{vn1}\be
\tilde{\zeta}(\tilde{x}_{1},\tilde{x}_{2})=\sum_{l=0}^{\infty}k_{l}
v_{l}(\tilde{x}_{1})v_{l}(\tilde{x}_{2}),\label{3}\ee where $\langle
v_{k}|v_{l}\rangle=\delta_{kl}$. By changing the variables back in
(\ref{3}), namely, by $ \tilde{x}_{1}\mapsto x_{1}+{x_{c}/
\sqrt{2}}, \tilde{x}_{2}\mapsto x_{2}-{x_{c}/ \sqrt{2}}$, one
arrives at $\zeta(x_{1},x_{2})=\sum_{l=0}^{\infty}k_{l}
\mbox{L}_{l}(x_{1})\mbox{R}_{l}(x_{2})$, where the orbitals $
\mbox{L}_{l}(x)=v_{l}(x+{x_{c}/ \sqrt{2}});
\mbox{R}_{l}(x)=v_{l}(x-{x_{c}/ \sqrt{2}})$ satisfy $\langle
\mbox{L}_{k}|\mbox{L}_{l}\rangle=\langle
\mbox{R}_{k}|\mbox{R}_{l}\rangle=\delta_{kl}$. Finally, after
substituting the expansions of $\zeta(x_{1},x_{2})$ and of
$\zeta(x_{2},x_{1})$ into (\ref{ex}), we get
\begin{equation} \psi^{(\pm)}(x_{1},x_{2})=\nonumber\\\sum_{l=0}^{\infty}k_{l}
[\mbox{L}_{l}(x_{1})\mbox{R}_{l}(x_{2})\pm
\mbox{R}_{l}(x_{1})\mbox{L}_{l}(x_{2})],\label{sh}\end{equation} It
is easy to infer  that  in  the limit as $g\rightarrow\infty$, where
$x_{c}\rightarrow\infty$, the integral overlap between the orbitals
$\mbox{L}_{k}(x)$ and $\mbox{R}_{l}(x)$ vanishes for any $k,l$,
$\langle \mbox{L}_{k}|\mbox{R}_{l}\rangle=0$. Hence, bearing in mind
that $\langle \mbox{L}_{k}|\mbox{L}_{l}\rangle=\langle
\mbox{R}_{k}|\mbox{R}_{l}\rangle=\delta_{kl}$,  one can    infer
  that  the family
$\{\mbox{L}_{l}(x)$, $\mbox{R}_{l}(x)\}$ forms asymptotically an
orthonormal set as $g\rightarrow\infty$. In this limit the orbitals
$\mbox{L}_{l}(x)$ and $\mbox{R}_{l}(x)$  are thus  nothing else but
the asymptotic natural orbitals  with the occupancy $\lambda_{l}$
related to the corresponding coefficient $k_{l}$ by
$\lambda_{l}=k_{l}^2$ (a two-fold degeneracy occurs). The bosonic
and fermionic ground states have thus as $g\rightarrow\infty$ the
same set of  occupancies. Up to this point, our derivations are
quite similar to those in \cite{9.1}, where the system of two
Coulombically interacting particles was treated by the HA. Here we
 go
 one step further and derive   closed form analytical expressions  for the asymptotic occupancies and their natural
 orbitals basing on a methodology of
    \cite{nagy}, wherein a
novel derivation of the occupancies  of the analytically solvable
two-particle Moshinsky model was given.

Following \cite{nagy},  we start with
 Mehler's formula: \begin{eqnarray}e^{-
(u^2+v^2){z^2\over 1-z^2} +uv {2z\over 1-z^2}}
=\nonumber\\=\sum_{l=0}^{\infty}\sqrt{1-z^2}({z\over 2})^l
{\mbox{H}(l;u) \mbox{H}(l;v)\over l!},\label{meh}\end{eqnarray}
where $\mbox{H}(l;.)$ is the $l^{th}$ order Hermite polynomial. Now
it should be  clear that  the orbitals $v_{l}$ and their
coefficients $k_{l}$ appearing in  Eq. (\ref{3})
 can be found in closed forms by matching Eq. (\ref{polo}) with  (\ref{meh}). Indeed, in the
 limit as $g\rightarrow\infty$ that we are interested in, only
($C^{(\pm)}(g)\rightarrow
 {5^{1\over 8}/
\sqrt{2}{\pi}^{1\over 4} }$ in Eq. (\ref{polo})),  we arrive, after
performing some tedious algebra, at
$$k_{l}^{g\rightarrow\infty}={\sqrt{{1-z^2}\over 2}z^l},$$
and $$v_{l}^{g\rightarrow\infty}(\tilde{x})={w^{1\over 4}\over
\pi^{1\over 4}\sqrt{2^l l!}}e^{-{1\over 2}w
\tilde{x}^2}\mbox{H}(l;\sqrt{w} \tilde{x}),$$ where $z={(w-1)/(
w+1)}$, $w=5^{1/ 4}$.

Next, using  the analytical formula obtained above for
$\lambda_{l}^{g\rightarrow\infty}$,
$\lambda_{l}^{g\rightarrow\infty}=[k_{l}^{g\rightarrow\infty}]^2$,
we compute  the R\'{e}nyi entropy \cite{reny}, that is, \be
\mbox{S}_{R}^{g\rightarrow\infty}= {1\over
1-q}\mbox{log}_{2}(2\sum_{l=0}^{\infty}
[\lambda_{l}^{^{g\rightarrow\infty}}]^{q}),\label{jklpg}\ee  where
the factor $2$ comes from the normalization condition
($2\sum_{l=0}\lambda_{l}^{g\rightarrow\infty}=1$). It is easy to
check that in the limit as $q\rightarrow 1$, Eq. (\ref{jklpg})
reduces to
$\mbox{S}_{B}^{g\rightarrow\infty}=-2\sum_{l}\lambda_{l}^{^{g\rightarrow\infty}}
\mbox{log}_{2}\lambda_{l}^{^{g\rightarrow\infty}}$. By performing
the summation in (\ref{jklpg}), we get \be
\mbox{S}_{R}^{g\rightarrow\infty} ={1\over
1-q}\mbox{log}_{2}({2^{1-q}(1-z^2)^q\over 1-z^{2q} })\label{reny},
\ee and then,
 by taking the limit as $q\rightarrow1$, we arrive at
 \be \mbox{S}^{g\rightarrow\infty}_{B}={z^2\over
z^2-1}\mbox{log}_{2}z^2-\mbox{log}_{2}(1-z^2)+1,\label{cv}\ee
$\mbox{S}^{g\rightarrow\infty}_{B}\simeq1.24939$.  As discussed in
Section \ref{section2}, the asymptotic fermionic and bosonic ground
states share   the same set of occupancies, so that the values of
their vN entropies differ  from each other only by one,
$\mbox{S}^{g\rightarrow\infty}_{F}\simeq0.24939$. In the next
subsection we will confirm the correctness of our analytical
 results  by comparing them with the results obtained  numerically.


\subsection{Numerical analysis}\label{numk}

As far as we know, except for the cases discussed in the previous
subsections, neither analytical solutions to Eq. (\ref{EOM1}) nor to
Eq. (\ref{EOM2}) are known and   we have to resort to numerical
methods.  Here we apply  the
 Rayleigh--Ritz (RR)
method, that uses as a variational trial function
 a linear combination of a finite set
of  some basis functions,
\begin{equation}\phi(x)=\sum_{n=0}^{N-1} c_{n}u_{n}(x).\label{dddd}\end{equation}
Since the interaction  potential in Eq. (\ref{EOM1}),
$U(\sqrt{2}|x|)$, is finite at the origin $x=0$,
  the basis of  the harmonic oscillator (HO)
eigenfunctions appears to be appropriate in this case. We found it
to work well over the whole  range of values of $\epsilon$ and $g$.
On the other hand, when it comes to Eq. (\ref{EOM2}), the
interaction potential in it, $\sim|x|^{-3}$, diverges at $x=0$, and
the
 interaction integrals are not convergent in the HO
basis. Fortunately, this divergence can be cured  by using
   the pseudoharmonic oscillator basis \cite{hall}
\begin{equation} u_{n}(x)\sim
x^{\gamma-{1\over 2}}e^{-{1\over 2} x^2
}{_{1}\mbox{F}_{1}}(-n;\gamma; x^2), \label{Lag}\end{equation}
$0<x<\infty$, $(u_{n}(0)=0)$, where $_{1}\mbox{F}_{1}$ is the Kummer
confluent hypergeometric function and  $\gamma$ is a nonlinear
parameter ($\gamma>{3/ 2}$), which, as we have verified, has a
strong impact on the rate of the convergence of the RR method.

 We determine  the value of the
parameter $\gamma$ according to
  an optimization strategy  based on   the stationarity of the trace of the RR matrix, $H^{RR}$ \cite{trace},\begin{equation}
\frac{d}{d\gamma}\mbox{Tr}_{N}H^{RR}|_{\gamma=\gamma_{opt}}=0.\label{opty2}\end{equation}
Having the RR wavefunctions $\phi_{i}(x)$ obtained in that way
(where $\phi_{i}=0$ for $x<0$), we  can construct the even $(+)$ and
odd $(-)$ approximate solutions of (\ref{EOM2}) by putting
$\psi^{(\pm)}_{i}\sim\phi_{i}(x)\pm\phi_{i}(-x)$.

For the demonstration of the convergence, Table \ref{even0one} shows
the numerical results for  the ground-state energy of (\ref{EOM2})
together with the corresponding optimal values of $\gamma_{opt}$,
for different values of  $g$ and $N$. As can be seen, the
convergence is fairly good over  the full interacting regime,
getting increasingly better   with an increase in  $g$. It is
apparent from the results that the occurrence of  the
 TG regime, which is
manifested by the closeness of the relative energy
 value  to $1.5$, takes place at a very small value of $g$, that is, at about $g=
0.0001$.

\begin{table}[h]
  \centering
  \begin{tabular}{cccc}
    \hline
 $g$  & $N$ & $\gamma_{opt} $& $E_{0}^{(N)}(\gamma_{opt})$    \\
    \hline
$0.0001$& 40  &1.577 &1.50333\\

& 50  &1.581 &1.50330\\

\hline
$0.01$& 40  &1.940&1.58259\\

& 50  &1.960&1.58249\\
\hline

$1$ & 30  &4.092 &2.67084 \\
& 40  &4.228 & 2.67079\\

\hline

$5$   & 20  & 5.942&3.98386 \\
   & 30  &6.240 &3.98383 \\
\hline
$1000$& 5  & 30.37& 24.6666\\
   & 10  &31.29 &24.6665 \\

    \hline
  \end{tabular}
   \caption{The lowest  approximate energy $E_{0}^{(N)}(\gamma_{opt})$  determined using the optimized RR
   method as discussed in the text.
}\label{even0one}
\end{table}


\begin{figure}[h]
\begin{center}
\includegraphics[width=0.45\textwidth]{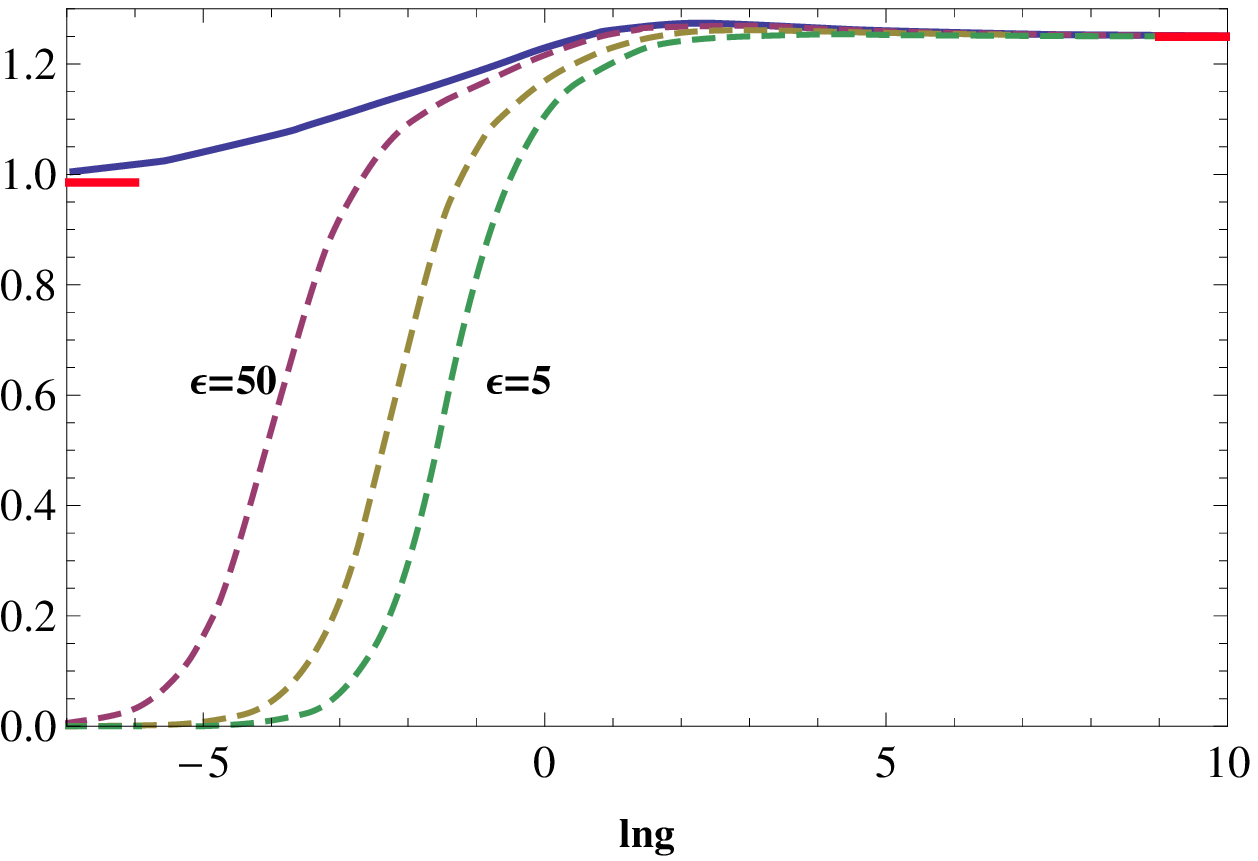}

\includegraphics[width=0.45\textwidth]{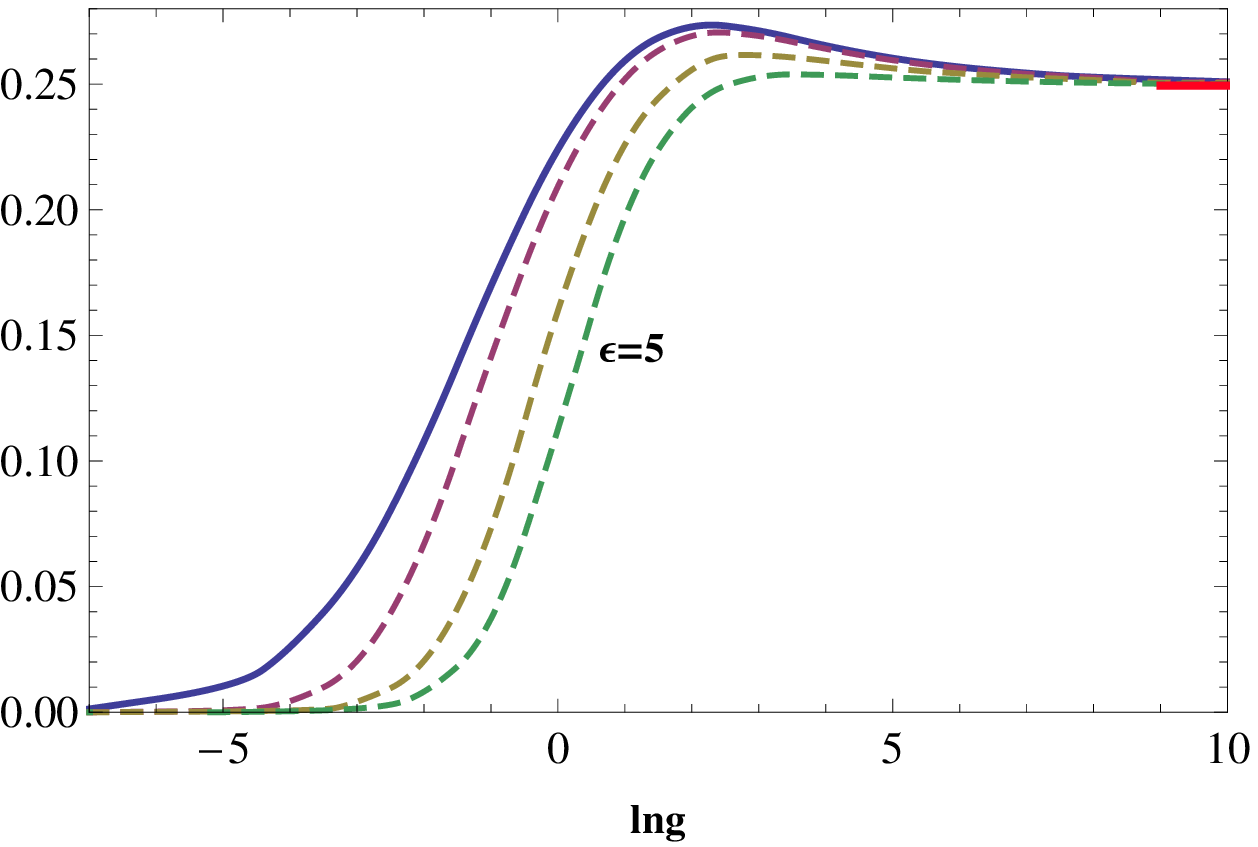}

\end{center}
\caption{Upper figure: The dependence of the vN entropy of the
bosonic ground-state on $\ln g$ for  $\epsilon=5,10,50$ (dashed
lines) and $\epsilon\rightarrow\infty$ (continuous line). The
asymptotic values are marked by horizontal lines. Lower figure: Same
as in upper figure but only for fermions.} \label{fig:od3og1}
\end{figure}

  Fig. \ref{fig:od3og1}
depicts our numerical results for the vN entropies of the bosonic
and fermionic ground states calculated for some exemplary values of
$\epsilon$, including the limiting case of
$\epsilon\rightarrow\infty$. The numerical results for
$\mbox{S}^{g\rightarrow0}_{B}$ and
 the analytical ones for $\mbox{S}^{g\rightarrow\infty}_{B}$ and
for $\mbox{S}^{g\rightarrow\infty}_{F}$  obtained  in the previous
subsections
 are marked by horizontal lines. As may be seen,  the vN entropy increases, attains a local maximum,
and then decreases, and in the limit of large $g$
 saturates  at a  value that is insensitive to
$\epsilon$. As a matter of fact, the last point has been already
explained at the beginning of the subsection
 \ref{dd}. As the figure indicates, our analytical results concerning
the limit $g\rightarrow\infty$ are in excellent agreement with the
numerical ones, which confirms their correctness and thereby proves
the applicability of the HA to the case of dipolar particles.
 The dependence of the entanglement on $\epsilon$ is
 stronger in the bosonic case than in the fermionic one, which
can be attributed to the fact that
 the probability
of finding the bosons at the same place is, in contrast to fermions,
affected by $\epsilon$ in particular.
 This effect is less
pronounced at large $g$, where the bosons become spatially separated
 simulating thus the Pauli exclusion principle. Both in the bosonic and fermionic case, the vN entropy is
  largest in the 1D limit and its value  attained in this limit
by  a local maximum is maximal. We see the TG regime starting   to
occur at a very small value of $g$, which is consistent with our
earlier conclusion drawn from the  numerical results  for the
relative motion energy (Table \ref{even0one}). When it comes to
 the points
 appearing  in the behaviour of the bosonic vN
entropy at which $\mbox{S}_{B}=1$, we have verified  their
$\mbox{Sn}$ are essentially different from $2$, so they
 must be considered as truly entangled.

We close our discussion with a comparison  of the present results
obtained  for the systems with the DDI  with the ones obtained in
\cite{kos10} for quasi 1D systems of two harmonically trapped
particles with a  Coulomb  interaction. Here we refer only to the
strong correlation regimes of both systems, comparing them regarding
their entanglement features. In particular, in \cite{kos10}, the
value of the vN entropy of strongly interacting charged bosons
 was found  to be about
 $1.14$.   Comparing this value with that
obtained in this paper,
$\mbox{S}^{g\rightarrow\infty}_{B}\approx1.25$, leads us to the
conclusion that  strongly interacting  dipolar bosons are more
entangled than  the strongly interacting charged ones.

\section{Summary}\label{sum}

We carried out, within the single mode approximation (SMA), a
comprehensive study of the entanglement properties of systems of two
dipolar particles confined in  a harmonic trap. Our results show the
effect of the anisotropy parameter $\epsilon$ on the entanglement in
the bosonic and fermionic ground states over the whole range of
values of $g$. Moreover, within the
  framework of the harmonic approximation (HA), closed-form  analytical
expressions for the  asymptotic natural orbitals and their
occupancies have been obtained by use of   Mehler's formula. The
corresponding  vN entropies of the asymptotic  bosonic and fermionic
ground states
  have been also derived analytically
and it has been shown that their values are in excellent agreement  with the results obtained numerically. 

\bibliography{aipsamp}

\end{document}